\RequirePackage{fix-cm}

\documentclass[onecolumn,epjc3]{svjour3}  
\makeatletter
\def\cl@chapter{\@elt {theorem}}
\makeatother

\smartqed  
\RequirePackage{graphicx}
\usepackage{dcolumn}
\usepackage{verbatim}

\usepackage{breakurl}
\usepackage{lmodern}

\usepackage{graphicx}
\usepackage{pgffor}
\usepackage{bm}
\usepackage{amsmath}
\usepackage{amsfonts}
\usepackage{multirow}
\usepackage[colorlinks,linktocpage,linkcolor=cyan,citecolor=cyan]{hyperref}
\usepackage{adjustbox}
\usepackage{array}
\usepackage[capitalise]{cleveref}
\usepackage{acro}
\usepackage[utf8]{inputenc}
\usepackage{CJKutf8}
\usepackage{amssymb}
\usepackage[justification=raggedright,singlelinecheck=false]{caption}
\usepackage{subcaption}

\usepackage{tikz}
\usepackage{pgfplots}
\pgfplotsset{compat=newest,every axis plot/.append style={line width=1pt}}

\crefname{figure}{Fig.}{Figs.}
\Crefname{figure}{Fig.}{Figs.}
\def\({\left(}
\def\){\right)}
\def\[{\left[}
\def\]{\right]}

\newcommand{\be}{{\begin{eqnarray}}}
\newcommand{\ee}{{\end{eqnarray}}}

\newcommand{\fnl}{f_\mathrm{NL}}

\newcommand{\Beq}{\begin{align}}
\newcommand{\Eeq}{\end{align}}

\DeclareAcronym{SW}{
  short = SW ,
  long = Sachs-Wolfe ,
  short-plural =  ,
}
\DeclareAcronym{ISW}{
  short = ISW ,
  long = integrated Sachs-Wolfe ,
  short-plural =  ,
}
\DeclareAcronym{BH}{
  short = BH ,
  long = black hole ,
  short-plural = s ,
}
\DeclareAcronym{SNR}{
  short = SNR ,
  long = signal-to-noise ratio ,
  short-plural = s ,
}
\DeclareAcronym{IMRPPv2}{
  short = ,
  long = {\normalsize IMRP}{\footnotesize HENOM}{\normalsize P}v2 ,
  short-plural = ,
}
\DeclareAcronym{SFR}{
  short = SFR ,
  long = star formation rate ,
  short-plural =  ,
}
\DeclareAcronym{IMR}{
  short = IMR ,
  long = inspiral-merger-ringdown ,
  short-plural =  ,
}
\DeclareAcronym{ABH}{
	short = ABH ,
	long  = astrophysical black hole,
  short-plural = s ,
}
\DeclareAcronym{GW}{
  short = GW ,
  long = gravitational wave ,
  short-plural = s ,
}
\DeclareAcronym{SIGW}{
  short = SIGW ,
  long = scalar-induced gravitational wave ,
  short-plural = s ,
}
\DeclareAcronym{GWB}{
  short = GWB ,
  long = gravitational-wave background ,
  short-plural = s ,
}
\DeclareAcronym{CBC}{
  short = CBC ,
  long = compact binary coalescence ,
  short-plural = s ,
}
\DeclareAcronym{BBH}{
  short = BBH ,
  long = binary black hole ,
  short-plural = s ,
}
\DeclareAcronym{PBH}{
  short = PBH ,
  long = primordial black hole ,
  short-plural = s ,
}
\DeclareAcronym{LIGO}{
  short =LIGO ,
  long = Laser Interferometer Gravitational-Wave Observatory ,
  short-plural = ,
}
\DeclareAcronym{LVK}{
  short = LVK ,
  long = {Advanced LIGO, Virgo and KAGRA} ,
  short-plural = ,
}
\DeclareAcronym{ET}{
	short = ET ,
	long  = Einstein Telescope, 
  short-plural =  ,
}
\DeclareAcronym{CE}{
	short = CE ,
	long  = Cosmic Explorer, 
  short-plural =  ,
}
\DeclareAcronym{LISA}{
	short = LISA ,
	long  = Laser Interferometer Space Antenna,
  short-plural =  ,
}
\DeclareAcronym{BBO}{
	short = BBO ,
	long  = Big Bang Observer,
  short-plural =  ,
}
\DeclareAcronym{DECIGO}{
	short = DECIGO ,
	long  = Deci-hertz Interferometer Gravitational wave Observatory,
  short-plural =  ,
}
\DeclareAcronym{SKA}{
	short = SKA ,
	long  = Square Kilometre Array,
  short-plural =  ,
}
\DeclareAcronym{PTA}{
  short = PTA ,
  long = pulsar timing array ,
  short-plural = s ,
}
\DeclareAcronym{NANOGrav}{
	short = NANOGrav ,
	long  = North American Nanohertz Observatory for Gravitational Waves , 
  short-plural =  ,
}
\DeclareAcronym{FRW}{
  short = FRW ,
  long = Friedman-Robertson-Walker ,
  short-plural =  ,
}
\DeclareAcronym{CMB}{
  short = CMB ,
  long = cosmic microwave background ,
  short-plural =  ,
}
\DeclareAcronym{LSS}{
  short = LSS ,
  long = large-scale structures ,
  short-plural =  ,
}
\DeclareAcronym{RD}{
  short = RD,
  long  = radiation-dominated ,
  short-plural =  ,
}
\DeclareAcronym{MD}{
  short = MD,
  long  = matter-dominated ,
  short-plural =  ,
}
\DeclareAcronym{HD}{
  short = HD,
  long  = Hellings-Downs ,
  short-plural =  ,
}
\DeclareAcronym{SMBH}{
  short = SMBH ,
  long  = supper-massive black hole ,
  short-plural = s ,
}
\DeclareAcronym{SGWB}{
  short = SGWB ,
  long  = stochastic gravitational-wave background ,
  short-plural = s ,
}
\DeclareAcronym{NG15}{
  short = NG15 ,
  long  = NANOGrav 15-year ,
  short-plural =  ,
}
\DeclareAcronym{PSD}{
  short = PSD ,
  long  = power spectral density ,
  short-plural = s ,
}
\DeclareAcronym{PDF}{
  short = PDF ,
  long  = probability distribution function ,
  short-plural = s ,
}
\DeclareAcronym{CV}{
  short = CV ,
  long  = cosmic variance ,
  short-plural =  ,
}

\journalname{Eur. Phys. J. C}
\begin{document}

\title{Study of primordial non-Gaussianity $f_{\mathrm{NL}}$ and $g_{\mathrm{NL}}$ with the cross-correlations between the scalar-induced gravitational waves and the cosmic microwave background}


\author{Zhi-Chao Zhao\thanksref{addr1}
        \and
        Sai Wang\thanksref{e1,addr2} 
        \and
        Jun-Peng Li\thanksref{addr3,addr4} 
        \and
        Kazunori Kohri\thanksref{addr5,addr6,addr7} 
}

\thankstext{e1}{Correspondence author: wangsai@ihep.ac.cn}


\institute{Department of Applied Physics, College of Science, China Agricultural University, 17 Qinghua East Road, Haidian District, Beijing 100083, China \label{addr1}
           \and
           School of Physics, Hangzhou Normal University, No.2318 Yuhangtang Road, Yuhang District, Hangzhou 311121, China \label{addr2}
           \and
           Theoretical Physics Division, Institute of High Energy Physics, Chinese Academy of Sciences, 19B Yuquan Road, Shijingshan District, Beijing 100049, China \label{addr3}
           \and
           School of Physics, University of Chinese Academy of Sciences, 19A Yuquan Road, Shijingshan District, Beijing 100049, China \label{addr4}
           \and
           Division of Science, National Astronomical Observatory of Japan (NAOJ), and SOKENDAI, 2-21-1 Osawa, Mitaka, Tokyo 181-8588, Japan \label{addr5}
           \and
           Theory Center, IPNS, and QUP (WPI), KEK, 1-1 Oho, Tsukuba, Ibaraki 305-0801, Japan \label{addr6}
           \and
           Kavli IPMU (WPI), UTIAS, The University of Tokyo, Kashiwa, Chiba 277-8583, Japan \label{addr7}
}

\date{Received: date / Accepted: date}

\maketitle

\begin{abstract}
The stochastic gravitational-wave background originating from cosmic sources contains vital information about the early universe. In this work, we comprehensively study the cross-correlations between the energy-density anisotropies in scalar-induced gravitational waves (SIGWs) and the temperature anisotropies and polarization in the cosmic microwave background (CMB). In our analysis of the angular power spectra for these cross-correlations, we consider all contributions of the local-type primordial non-Gaussianity $f_{\mathrm{NL}}$ and $g_{\mathrm{NL}}$ that can lead to large anisotropies. We show that the angular power spectra are highly sensitive to primordial non-Gaussianity. Furthermore, we project the sensitivity of future gravitational-wave detectors to detect such signals and, consequently, measure the primordial non-Gaussianity.
\end{abstract}

\section{Introduction}\label{sec:introduction}

\Acp{SIGW} are generated nonlinearly by the linear cosmological curvature perturbations in the early universe \cite{Ananda:2006af,Baumann:2007zm,Espinosa:2018eve,Kohri:2018awv,Mollerach:2003nq,Assadullahi:2009jc,Domenech:2021ztg}, with their energy-density spectrum influenced by primordial non-Gaussianity \cite{Maldacena:2002vr,Bartolo:2004if,Allen:1987vq,Bartolo:2001cw,Acquaviva:2002ud,Bernardeau:2002jy,Chen:2006nt,Byrnes:2010em}. This non-Gaussianity reflects the level of interaction of the inflaton field. Additionally, the production of \acp{SIGW} may coincide with the creation of \acp{PBH}, which are considered viable candidates for dark matter \cite{Carr:2020xqk,Carr:2020gox,Carr:2023tpt}. Consequently, \acp{SIGW} serve as crucial sources of information regarding the universe's origin and evolution during a pre-recombination era that is unaccessible with other cosmological tracers, while also potentially serving as probes of dark matter. However, observations of \acp{SIGW} may face contamination from astrophysical foregrounds, such as a \acl{SGWB} from compact binary coalescences \cite{Regimbau:2011rp}. Moreover, the energy-density spectrum of \acp{SIGW} exhibits degeneracy in model parameters. To address these challenges, it is essential to introduce a comprehensive set of observables that characterize \acp{SIGW} effectively.

Studies have shown that the auto-correlations among the energy-density anisotropies in \acp{SIGW} can be a valuable tool for removing astrophysical foregrounds and addressing parameter degeneracy \cite{LISACosmologyWorkingGroup:2022kbp,Li:2023qua,Li:2023xtl}. It was shown that the presence of significant anisotropies in \acp{SIGW} is contingent upon the local-type primordial non-Gaussianity \cite{Bartolo:2019zvb,Li:2023qua,Li:2023xtl,Wang:2023ost,Yu:2023jrs,Ruiz:2024weh,Rey:2024giu}. This finding underscores the potential of the angular power spectrum as a key indicator of inflation dynamics and a means to differentiate between various inflation models. It was also shown that the angular bispectra and trispectra of \acp{SIGW} are direct probes of primordial non-Gaussianity \cite{Bartolo:2019zvb,Li:2024zwx}. However, the possible presence of parameter degeneracy, unknown systematics, and residual foregrounds in these auto-correlations highlights the need for additional observables to effectively address this challenge.

Cross-correlations between \acp{SIGW} and other cosmological tracers are being established to extract crucial information about the universe, getting rid of shortages of the auto-correlations. Recent analyses have investigated the cross-correlations between the energy-density anisotropies in \acp{SIGW} and the temperature anisotropies in the \ac{CMB} \cite{Dimastrogiovanni:2022eir,Schulze:2023ich,Cai:2024dya}. These cross-correlations can be significant due to the shared origin of both anisotropies and the common influence of large-scale density perturbations on the propagation of gravitons and photons. Future advancements in gravitational-wave detectors and \ac{CMB} experiments are poised to offer precise constraints on the non-Gaussian parameters through these cross-correlations \cite{Malhotra:2020ket,Perna:2023dgg,Adshead:2020bji}. Therefore, the integration of both auto- and cross-correlations would enhance our comprehension of the early universe.

In this study, we conduct a comprehensive investigation into the cross-correlations between \acp{SIGW} and the \ac{CMB}. Our analysis encompasses the influence of local-type primordial non-Gaussianity, parameterized as $f_{\mathrm{NL}}$ and $g_{\mathrm{NL}}$, on the energy density of \acp{SIGW} at both background and anisotropy levels \cite{Li:2023xtl}. We also examine the cross-correlations involving polarization in the \ac{CMB}, besides temperature anisotropies. Furthermore, we demonstrate that the parameter degeneracy can be resolved by taking into account these cross-correlations. Through the application of the Fisher-matrix method, we evaluate the sensitivity of future gravitational-wave detectors and \ac{CMB} experiments in detecting these signals and accurately measuring the non-Gaussian parameters.

The remaining of the paper is arranged as follows. 
By briefly summarizing the formulas for the energy-density anisotropies in \acp{SIGW}, we provide the formulas for the cross-correlations between \acp{SIGW} and the \ac{CMB} in Section~\ref{sec:cross-correlation}.
We perform the Fisher-matrix forecastings for the non-Gaussian parameter $\fnl$ in Section~\ref{sec:fisher}.
We demonstrate the conclusions and discussion in Section~\ref{sec:conclusion}.

\section{Cross-correlations between SIGWs and the CMB}\label{sec:cross-correlation}

In this section, we provide a brief overview of the fundamental formulas, with more comprehensive details available in Ref.~\cite{Li:2023xtl}.

\subsection{Temperature anisotropies and polarization in the CMB}

Considerable research has been dedicated to investigating temperature anisotropies and polarization in the \ac{CMB} \cite{Seljak:1996is,Zaldarriaga:1996xe,Kamionkowski:1996ks}. Photons from the early universe follow a thermal distribution resembling a blackbody spectrum. Inhomogeneities in the energy density of these relic photons result in temperature fluctuations, which are observed as temperature anisotropies in the \ac{CMB}. Additionally, the presence of a quadrupole moment gives rise to polarization signals within the \ac{CMB}. 
In this section, we present the formulas for temperature anisotropies (T) and polarization (E), excluding the consideration of B-mode polarization due to its lack of correlation with \acp{SIGW} resulting from parity constraints.

The standard practice is to express the T and E modes of the \ac{CMB} in harmonic space, where the coefficients of spherical harmonics are denoted as
\begin{equation}
a_{X \ell m} = 4\pi(-i)^{\ell} \int \frac{d^{3}\bold{k}}{(2\pi)^{3}} e^{i\bold{k}\cdot\bold{x}_{0}} Y_{\ell m}^{\ast}(\hat{\bold{k}}) \mathcal{R}(\bold{k}) \Delta_{X \ell}(k,\eta_{0})\ ,
\end{equation}
where $\mathcal{R}$ represents the primordial curvature perturbations, $\bold{k}$ denotes the wavevector of these perturbations with $\hat{\bold{k}}$ indicating its direction, and $\Delta_{X \ell}$ corresponds to the transfer function of X (either T or E). Typically, the determination of $\Delta_{X \ell}$ involves solving the Boltzmann-Einstein equations, a task often performed using the publicly accessible \texttt{CLASS} \footnote{\url{https://github.com/lesgourg/class\_public}} software \cite{Diego_Blas_2011}.

The angular power spectrum of the \ac{CMB} is defined as 
\begin{equation}
\langle a_{X \ell m} a^{\ast}_{X' \ell' m'} \rangle = \delta_{\ell \ell'}\delta_{mm'}C^{\mathrm{XX'}}_{\ell}\ , 
\end{equation}
for which we have assumed the cosmological principle. 
It is explicitly shown as 
\begin{equation}
C^{\mathrm{XX'}}_{\ell} = 4\pi \int d\ln k\, \Delta_{X}(k,\eta_{0}) \Delta_{X'}(k,\eta_{0}) P_{L}(k) \ ,
\end{equation}
where $P_{L}$ represents the dimensionless power spectrum of $\mathcal{R}$ at large scales. 
For simplicity,  we assume that the spectral tilt vanishes, i.e., \cite{Jiang:2022uyg,Ye:2022efx} 
\begin{equation}
P_{L}(k) = A_{L} \ ,
\end{equation} 
where $A_{L}$ represents the spectral amplitude. 
However, it can be extended to accommodate other spectral tilts if necessary.
In this study, the subscript $_{0}$ denotes cosmological quantities in the present-day cosmos, with detailed values provided in Ref.~\cite{Planck:2018vyg}.

\subsection{Average background and the energy-density anisotropies in SIGWs}

In contrast to relic photons, gravitons are intrinsically non-thermal. It is crucial to comprehend the energy-density spectrum of \acp{SIGW} before delving into the incorporation of energy-density fluctuations, which can manifest as energy-density anisotropies within \acp{SIGW}. While gravitons and relic photons travel along similar geodesics, the origins of energy-density anisotropies in \acp{SIGW} do not perfectly align with temperature anisotropies in the \ac{CMB}. This discrepancy stems from the nonlinear production of \acp{SIGW} by linear cosmological curvature perturbations, necessitating the consideration of non-adiabatic initial conditions instead of the adiabatic conditions typically assumed for the \ac{CMB}. Such initial conditions can lead to significant anisotropies in \acp{SIGW} in principle.

\subsubsection{Average background}

The energy-density (fraction) spectrum of \acp{SIGW} at the average background level has been comprehensively investigated in Refs.~\cite{Li:2023xtl,Yuan:2023ofl,Ruiz:2024weh} by considering the local-type primordial non-Gaussianity parameters $f_{\mathrm{NL}}$ and $g_{\mathrm{NL}}$. {\color{black}In this work, by saying “average background”, we mean the monopole, i.e., isotropic component of the SGWB, which is denoted by $\bar{\Omega}_{\mathrm{gw}}$.} Additional studies on this topic are available in Refs.~\cite{Li:2023qua,Wang:2023ost,Nakama:2016gzw,Garcia-Bellido:2017aan,Adshead:2021hnm,Ragavendra:2021qdu,Yu:2023jrs,Ragavendra:2020sop,Garcia-Saenz:2022tzu,Abe:2022xur,Cai:2018dig,Unal:2018yaa,Atal:2021jyo,Yuan:2020iwf,Zhang:2021rqs,Lin:2021vwc,Chen:2022dqr,Chang:2023aba,Cai:2019elf,Ragavendra:2020sop,Garcia-Saenz:2022tzu,Yu:2023jrs,Perna:2024ehx,Zeng:2024ovg}. In our approach, we utilize diagrammatic techniques to derive semi-analytical formulas for this spectrum, as detailed in Ref.~\cite{Li:2023xtl}.
Specifically, the energy-density spectrum is expressed as
\begin{eqnarray}
\Omega_{\mathrm{gw}}(\eta_{i},q) &=& \Omega_{\mathrm{gw}}^{(0,0)} + \Omega_{\mathrm{gw}}^{(0,1)} + \Omega_{\mathrm{gw}}^{(1,0)} +\Omega_{\mathrm{gw}}^{(0,2)} +\Omega_{\mathrm{gw}}^{(1,1)}  \nonumber\\
&& + \Omega_{\mathrm{gw}}^{(2,0)} +\Omega_{\mathrm{gw}}^{(0,3)} +\Omega_{\mathrm{gw}}^{(1,2)} +\Omega_{\mathrm{gw}}^{(0,4)} \ ,
\end{eqnarray}
where $\eta_{i}$ signifies the production time of \acp{SIGW}, $q$ is the gravitational-wave wavenumber, and {\color{black} the expressions for $\Omega_{\mathrm{gw}}^{(a,b)}$ are sufficiently complex to make listing them in this paper impractical, but their semi-analytic results can be found in Fig.~5 of Ref.~\cite{Li:2023xtl}.}
Regarding perturbations that produce \acp{SIGW}, we consider the dimensionless power spectrum of $\mathcal{R}$ at small scales to be
\begin{equation}
P_{S}(k) = \frac{A_{S}}{\sqrt{2\pi \sigma^{2}}} \mathrm{exp}\left[ -\frac{\ln^{2}(k/k_{p})}{2\sigma^{2}} \right] \ ,
\end{equation}
where $A_{S}$ and $\sigma$ represent the spectral amplitude and width, with $k_{p}=2\pi f_{p}$ denoting the peak wavenumber with $f_{p}$ being the peak frequency. 
Assuming $\sigma=1$, we graphically represented $\Omega_{\mathrm{gw}}^{(a,b)}$ as a function of $q$ in Fig.~5 of Ref.~\cite{Li:2023xtl} for clarity.

Ultimately, we derive the energy-density spectrum of \acp{SIGW} in the present-day universe as
\begin{equation}
\Omega_{\mathrm{gw},0}(\nu) \simeq \Omega_{r,0}  \Omega_{\mathrm{gw}}(\eta_{i},q) \ ,
\end{equation}
where $\Omega_{r,0}=4.2\times10^{-5}h^{-2}$, with $h$ representing the dimensionless Hubble constant, denotes the energy-density fraction of radiation in the present-day universe, and $\nu=q/(2\pi)$ is the gravitational-wave frequency.
In the above equation, we neglect contributions from effective relativistic species during the thermal evolution of the early universe, which can be reintroduced if required.
Assuming $\sigma=1$, we presented illustrations of $\Omega_{\mathrm{gw},0}$ as a function of $\nu$ in Fig.~6 of Ref.~\cite{Li:2023xtl} for clarity.

\subsubsection{Energy-density anisotropies}

The investigation of energy-density anisotropies in \acp{SIGW} has recently been explored in Refs.~\cite{Bartolo:2019zvb,Li:2023qua,ValbusaDallArmi:2020ifo,Dimastrogiovanni:2021mfs,LISACosmologyWorkingGroup:2022kbp,LISACosmologyWorkingGroup:2022jok,Unal:2020mts,Malhotra:2020ket,Malhotra:2022ply,ValbusaDallArmi:2023nqn,Wang:2023ost,Li:2023xtl,Yu:2023jrs,Li:2024zwx,Schulze:2023ich,Rey:2024giu}.
Specifically, we presented all contributions of $f_{\mathrm{NL}}$ and $g_{\mathrm{NL}}$ to these anisotropies in Refs.~\cite{Li:2023xtl}.
Similar to the \ac{CMB}, the energy-density fluctuations of \acp{SIGW} can also be described in harmonic space.
The coefficients of spherical harmonics are provided as 
\begin{equation}
\delta_{\ell m} = 4\pi (-i)^{\ell} \int \frac{d^{3}\bold{k}}{(2\pi)^{3}} e^{i\bold{k}\cdot\bold{x}_{0}} Y_{\ell m}^{\ast}(\hat{\bold{k}}) \mathcal{R}(\bold{k}) \mathcal{T}_{\ell}(q,k,\eta_{0})\ .
\end{equation}
Here, the transfer function $\mathcal{T}_{\ell}$ can be derived by solving the Boltzmann-Einstein equations \cite{Contaldi:2016koz,Bartolo:2019zvb}. 
According to Ref.~\cite{Li:2023xtl}, we explicitly represent it as
\begin{eqnarray}\label{eq:tell}
\mathcal{T}_{\ell} = \frac{\Omega_{\mathrm{gw},0}}{4\pi} &\bigg\{& \left(\frac{\Omega_{\mathrm{ng}}}{\Omega_{\mathrm{gw}}} \right) j_{\ell}(k(\eta_{0}-\eta_{i}))  
+ \left(4-n_{\mathrm{gw}}\right)T_{\psi}(\eta_{i},k) j_{\ell}(k(\eta_{0}-\eta_{i})) \nonumber\\
&+& \left(4-n_{\mathrm{gw}}\right) \int_{\eta_{i}}^{\eta_{0}} d\eta \left(\frac{\partial T_{\phi}(\eta,k)}{\partial \eta}+\frac{\partial T_{\psi}(\eta,k)}{\partial \eta}\right) j_{\ell}(k(\eta_{0}-\eta)) \bigg\}
\end{eqnarray}
where the tilt of energy-density spectrum is defined as $n_{\mathrm{gw}}(\nu) = {d\ln \Omega_{\mathrm{gw},0}(\nu)}/{d\ln\nu}$, $\Omega_{\mathrm{ng}}$ denotes all terms involving the couplings of long- and short-wavelength perturbations, $T_{\psi}$ and $T_{\phi}$ represent the transfer functions of scalar perturbations in the comoving Newtonian gauge, and $j_{\ell}(x)$ is the spherical Bessel function. 
In this study, we utilize energy-density fluctuations instead of energy-density contrasts. The latter are defined as the former divided by the energy density per solid angle of the average background. 
This convention results in a factor of $\Omega_{\mathrm{gw},0}/(4\pi)$ in Eq.~(\ref{eq:tell}). 
We represent $\Omega_{\mathrm{ng}}$ as  
\begin{eqnarray}
\Omega_{\mathrm{ng}}(\eta_{i},q) &=& \frac{6f_{\mathrm{NL}}}{5} \left( 4\Omega_{\mathrm{gw}}^{(0,0)} +3\Omega_{\mathrm{gw}}^{(0,1)} +2\Omega_{\mathrm{gw}}^{(1,0)} +2\Omega_{\mathrm{gw}}^{(0,2)} +\Omega_{\mathrm{gw}}^{(1,1)} +\Omega_{\mathrm{gw}}^{(0,3)} \right)  \nonumber \\
&& + \frac{9g_{\mathrm{NL}}}{5f_{\mathrm{NL}}} \left( 2\Omega_{\mathrm{gw}}^{(1,0)} +2\Omega_{\mathrm{gw}}^{(1,1)} +4\Omega_{\mathrm{gw}}^{(2,0)} +2\Omega_{\mathrm{gw}}^{(1,2)} \right) \ ,
\end{eqnarray}
which aligns with Eq.~(3.14) in Ref.~\cite{Li:2023xtl}.
In Eq.~(\ref{eq:tell}), the first term on the right-hand side represents the initial inhomogeneities, while the second and third terms correspond to the \ac{SW} and \ac{ISW} effects \cite{Sachs:1967er}. 
The latter two effects are similar to those observed in the \ac{CMB}.

It is important to highlight that the non-adiabatic initial conditions differ from the adiabatic initial conditions assumed for the \ac{CMB}.
When we consider a large $|f_{\mathrm{NL}}|$ value, this distinct initial condition is responsible for the significant anisotropies in \acp{SIGW}.

\begin{figure}
    \centering
    \includegraphics[width=\linewidth]{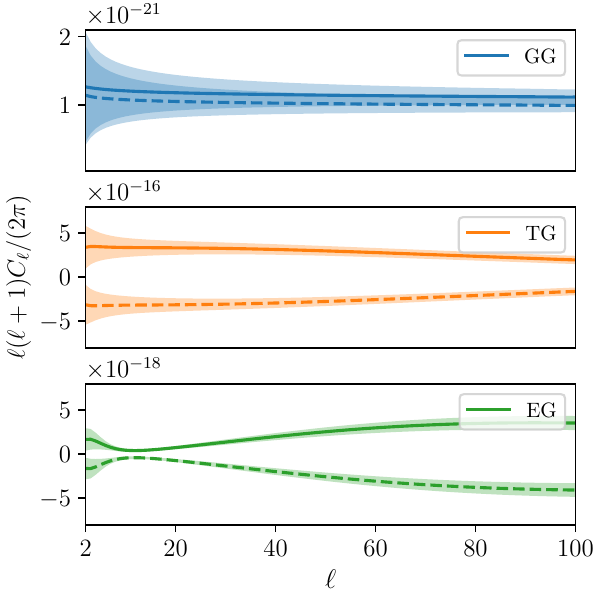}
    \caption{The angular power spectra for the auto- and cross-correlations are shown with \ac{CV} shaded. For illustration, we depict them in the frequency band of $\nu=f_{p}$ by assuming the model parameters $A_{S}=0.02$, $\sigma=1$, $f_{\mathrm{NL}}=\pm10$ (plus in solid while minus in dashed), and $g_{\mathrm{NL}}=20$. {\color{black}The shaded region shows the $\pm1\sigma$ (68\% C.L.) uncertainty from CV only.}}
    \label{fig:Fig1}
\end{figure}

\begin{figure}
    \centering
    \includegraphics[width=\linewidth]{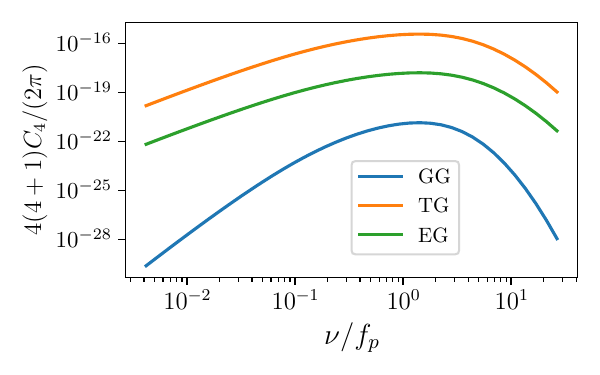}
    \caption{We show dependence of the angular power spectra on frequency bands. For the purpose of illustration, we fix $\ell=4$ and the model parameters $A_{S}=0.02$, $\sigma=1$, $f_{\mathrm{NL}}=10$, and $g_{\mathrm{NL}}=20$. }
    \label{fig:Fig2}
\end{figure}

In order to characterize the auto-correlations of energy-density anisotropies (G) in \acp{SIGW}, the angular power spectrum is defined as
\begin{equation}
\langle \delta_{\ell m} \delta_{\ell' m'}^{\ast} \rangle = \delta_{\ell\ell'}\delta_{mm'} C_{\ell}^{\mathrm{GG}}\ ,
\end{equation}
leading to the following expression
\begin{equation}\label{eq:cellgg}
C_{\ell}^{\mathrm{GG}} = 4\pi \int d\ln k\, \mathcal{T}_{\ell}^{2}(q,k,\eta_{0}) P_{L}(k) \ .
\end{equation}
In Fig.~\ref{fig:Fig1}, we present numerical results showcasing the auto-correlations of \acp{SIGW}, characterized by a flat angular power spectrum. 
{\color{black} We show the $1\sigma$ uncertainty due to \ac{CV}, namely $\Delta C_{\ell} = \sqrt{2/(2\ell +1)} C_{\ell}$, in shaded regions. }
Here, we focus on the auto-correlations within the same frequency band of gravitational waves. 
However, our research methodology can be readily extended to study correlations between different frequency bands, as demonstrated in Refs.~\cite{Li:2023qua,Li:2023xtl,Schulze:2023ich}. 
Based on Fig.~\ref{fig:Fig1}, we also illustrate that the auto-correlation spectrum exhibits significant degeneracy in model parameters, as discussed in  Refs.~\cite{Li:2023qua,Li:2023xtl}. 
In Fig.~\ref{fig:Fig2}, we demonstrate how auto-correlations vary with different frequency bands by considering a given angular multipole moment, i.e., $\ell=4$. 
Such dependence would be important to extracting vital information by performing the component separation \cite{Kuwahara:2024jiz,Liang:2024tgn}.

\subsection{Cross-correlations between SIGWs and the CMB}

The angular power spectrum for cross-correlations between \acp{SIGW} and the \ac{CMB} is defined as
\begin{equation}
\frac{1}{2}\langle a_{X\ell m} \delta_{\ell' m'}^{\ast} + a_{X \ell m}^{\ast} \delta_{\ell' m'} \rangle = \delta_{\ell \ell'} \delta_{m m'} C_{\ell}^{\mathrm{XG}}\ ,
\end{equation}
where the superscript $^{\mathrm{XG}}$ indicates the cross-correlations between X-mode of the \ac{CMB} and the energy-density anisotropies in \acp{SIGW}.
This is explicitly expressed as
\begin{equation}\label{eq:cellxg}
C_{\ell}^{\mathrm{XG}} = 4\pi \int d\ln k\, \Delta_{X \ell}(k,\eta_{0}) \mathcal{T}_{\ell}(q,k,\eta_{0}) P_{L}(k) \ ,
\end{equation}
which stands as one of the key outcomes of our current study.
This result will be used to estimate the projected sensitivity of future experiments to our model parameters in the subsequent section.
We also illustrate numerical results for the cross-correlations in Fig.~\ref{fig:Fig1} and their dependence on frequency bands in Fig.~\ref{fig:Fig2}. 
Our analysis reveals that the angular power spectra for cross-correlations exhibit a nearly flat profile, with certain features being introduced by the \ac{CMB}. 
Based on Fig.~\ref{fig:Fig1}, we observe that the cross-correlation spectra can alleviate the degeneracy in model parameters, potentially enabling the extraction of model information. {\color{black}That is, cross-correlation spectra can help distinguish between positive and negative values of $f_{\mathrm{NL}}$.}
In addition, we expect that they exhibit dependence on frequency bands. 
Based on our analysis outlined in this section, we expect that a combination of auto- and cross-correlations would be useful to subtracting systematics and foregrounds from observations by using, e.g., component separation \cite{Kuwahara:2024jiz,Liang:2024tgn}. 
We would leave such analysis to future works, as they are outside the scope of our current work.

\section{Fisher-matrix forecastings}\label{sec:fisher}

In this section, utilizing the Fisher-matrix approach \footnote{Ref.~\cite{Tian:2024ljg} introduced a Bayesian approach, which could be applied to our current study in principle. Nevertheless, for illustrative purposes, we anticipate that the Fisher-matrix approach will be adequate for our theoretical analysis.} \cite{ly2017tutorialfisherinformation}, we investigate the precision of measuring model parameters by detecting the cross-correlations between \acp{SIGW} and the \ac{CMB} using upcoming gravitational-wave detection.
For the purpose of illustration, our focus is solely on the non-linear parameter $f_{\mathrm{NL}}$, and thereby let $g_{\mathrm{NL}}=0$. 
Nevertheless, our research methodology can be readily extended to encompass other model parameters as required. 
Throughout this work, we do not consider potential sources of astrophysical foregrounds, which are left to be studied by future works.

\begin{figure}
    \centering
    \includegraphics[width=\linewidth]{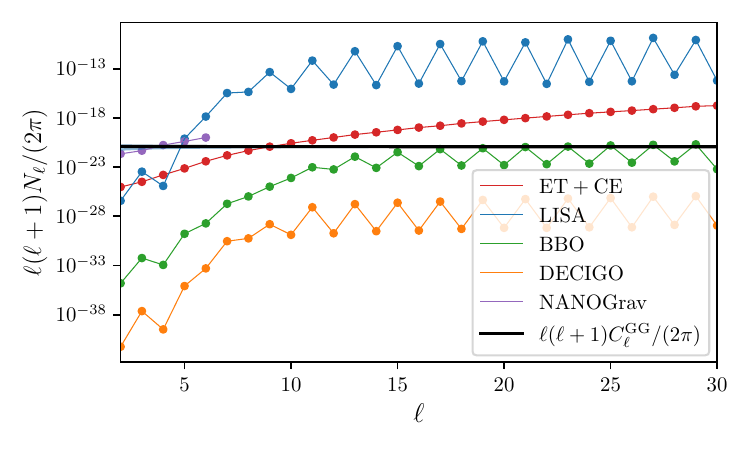}
    \caption{Comparison between the signal and the optimal sensitivity of detectors. The noise level of \ac{NANOGrav} is illustrated as the lowest in the lower panel of Fig.~4 in Ref.~\cite{Pol:2022sjn}, while the noise levels of other detectors are depicted in Fig.~7 of Ref.~\cite{Braglia:2021fxn}. In our analysis, we consider a set of model parameters $f_{\mathrm{NL}}=10$, $g_{\mathrm{NL}}=0$, $A_{S}=0.02$, $\sigma=1$, and $f_{p}=f_{\ast}$ with $f_{\ast}$ being the frequency associated with the optimal sensitivity. For the signal, we utilize the specified auto-correlated angular power spectrum at $\nu=f_{p}$. }
    \label{fig:Fig5}
\end{figure}

The \ac{SNR} for a combination of auto- and cross-correlations is defined as 
\begin{equation}
\mathrm{SNR}^{2} = \sum_{\ell = 2}^{\ell_{\max}} \bold{C}^{T} \mathbb{C}^{-1} \bold{C}\ ,\label{eq:sai02}
\end{equation}
where we define $\bold{C} = (C_{\ell}^{\mathrm{GG}},  C_{\ell}^{\mathrm{TG}}, C_{\ell}^{\mathrm{EG}})^{T}$, and the covariance matrix is represented as 
\footnote{Elements of $\mathbb{C}$ are specifically given as 
\begin{subequations}
\begin{align}
  {\sigma_{\ell}^{\mathrm{GG} - \mathrm{GG}}}^2 &= \frac{2}{2 \ell + 1} \left(C_{\ell}^{\mathrm{GG}} + N_{\ell}\right)^2  \ ,\\
  {\sigma_{\ell}^{\mathrm{GG} - \mathrm{TG}}}^2 &= \frac{2}{2 \ell + 1} \left[\left(C_{\ell}^{\mathrm{GG}} + N_{\ell}\right) C_{\ell}^{\mathrm{TG}}\right]  \ ,\\
  {\sigma_{\ell}^{\mathrm{GG} - \mathrm{EG}}}^2 &= \frac{2}{2 \ell + 1} \left[\left(C_{\ell}^{\mathrm{GG}} + N_{\ell}\right) C_{\ell}^{\mathrm{EG}}\right]  \ ,\\
  {\sigma_{\ell}^{\mathrm{TG} - \mathrm{TG}}}^2 &= \frac{1}{2 \ell + 1} \left[{\left({C_{\ell}^{\mathrm{TG}}}\right)^2 + C_{\ell}^{\mathrm{TT}} \left(C_{\ell}^{\mathrm{GG}} + N_{\ell}\right)}\right]  \ ,\\
  {\sigma_{\ell}^{\mathrm{TG} - \mathrm{EG}}}^2 &= \frac{1}{2 \ell + 1} \left[C_{\ell}^{\mathrm{TE}} \left(C_{\ell}^{\mathrm{GG}} + N_{\ell}\right) + C_{\ell}^{\mathrm{TG}} C_{\ell}^{\mathrm{EG}}\right]  \ ,\\
  {\sigma_{\ell}^{\mathrm{EG} - \mathrm{EG}}}^2 &= \frac{1}{2 \ell + 1}\left[{\left({C_{\ell}^{\mathrm{EG}}}\right)^2 + C_{\ell}^{\mathrm{EE}} \left(C_{\ell}^{\mathrm{GG}} + N_{\ell}\right)}\right]  \ .
\end{align}
\end{subequations}
}
\begin{equation}
  \mathbb{C} = \left( \begin{array}{ccc}
    {\sigma_{\ell}^{\mathrm{GG}
    - \mathrm{GG}}}^2 & {\sigma_{\ell}^{\mathrm{GG} - \mathrm{TG}}}^2 &
    {\sigma_{\ell}^{\mathrm{GG} - \mathrm{EG}}}^2\\
    {\sigma_{\ell}^{\mathrm{GG} - \mathrm{TG}}}^2 &
    {\sigma_{\ell}^{\textrm{TG$-$TG}}}^2 & {\sigma_{\ell}^{\mathrm{TG} -
    \mathrm{EG}}}^2\\
    {\sigma_{\ell}^{\mathrm{GG} - \mathrm{EG}}}^2 &
    {\sigma_{\ell}^{\mathrm{TG} - \mathrm{EG}}}^2 &
    {\sigma_{\ell}^{\textrm{EG$-$EG}}}^2
  \end{array} \right) \ .\label{eq:sai01}
\end{equation}
Our analysis includes the noise characteristics $N_{\ell}$ of gravitational-wave detectors \footnote{We use the \ac{NANOGrav} \cite{jenet2009northamericannanohertzobservatory}, \ac{LISA} \cite{Baker:2019nia,Smith:2019wny}, \ac{BBO} \cite{Crowder:2005nr,Smith:2016jqs}, \ac{DECIGO} \cite{Seto:2001qf,Kawamura:2020pcg}, and a network of \ac{ET} \cite{Hild:2008ng} and \ac{CE} \cite{Reitze:2019iox}. We utilize the noise from the most sensitive frequency band for each detector \cite{Pol:2022sjn,Braglia:2021fxn}. We depict them in Fig.~\ref{fig:Fig5} to compare with the specified signal. Similar methodologies can be applied to analyze other detectors.}, while considering only \ac{CV} for the \ac{CMB} since the noise in future \ac{CMB} experiments could potentially be dominated by \ac{CV}. 
When ignoring $N_{\ell}$, we obtain an inevitable uncertainty due to \ac{CV}.

{\color{black}To evaluate the precision of measuring model parameters, we utilize the Fisher matrix, denoted as $F$, for which detailed derivations have been explicitly shown in Refs.~\cite{Perna:2023dgg,Tegmark:1999ke}. Its components are given by 
\begin{equation}
        \begin{aligned} F_{i j} & =\sum_{\ell}^{\ell_{\max }} \frac{2 \ell+1}{2} \operatorname{Tr}\left[\frac{\partial C_{\ell}}{\partial \theta_i} C_{\ell}^{-1} \frac{\partial C_{\ell}}{\partial \theta_j} C_{\ell}^{-1}\right] \\ & =\sum_{\ell}^{\ell_{\max }} \frac{\partial \mathbf{C}^T}{\partial \theta_i} \mathbb{C}^{-1} \frac{\partial \mathbf{C}}{\partial \theta_j}\ ,\end{aligned}
        \end{equation}
where ${\partial \bold{C}}/{\partial \theta}$ denotes the derivative of each element of $\bold{C}$, as defined below Eq.~(\ref{eq:sai02}), with respect to a model parameter $\theta$, $\mathbb{C}$ is the covariance matrix, as defined in Eq.~(\ref{eq:sai01}), and $\ell_{\mathrm{max}}$ represents the maximum value of angular multipole moments that are relevant for a given detector.}
For the sake of clarity, we concentrate on a single parameter, specifically $p_{1} = f_{\mathrm{NL}}$.
The $1\sigma$ uncertainty of $f_{\mathrm{NL}}$ is determined by the Cramér-Rao bound, i.e., 
\begin{equation}
\sigma_{f_{\mathrm{NL}}} = \sqrt{\left(F^{-1}\right)_{11}} \ ,
\end{equation}
which provides a rough estimate of the measurement precision.
This level of analysis is adequate for our theoretical investigation.

\begin{figure}
    \centering
    \includegraphics[width=\linewidth]{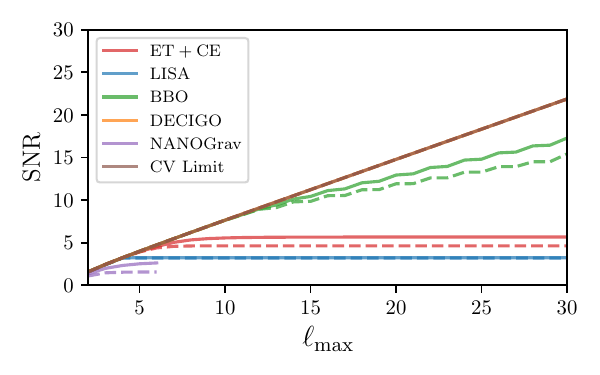}
    \caption{The \acp{SNR} are depicted for the detection of auto-correlations (dashed curves) and both auto- and cross-correlations (solid curves) using future gravitational-wave detectors. We adopt the same set of model parameters as those in Fig.~\ref{fig:Fig5}. For comparison, we include the \ac{CV} limits. It is worth noting that the curves corresponding to \ac{DECIGO} perfectly align with the \ac{CV} limits. }
    \label{fig:Fig3}
\end{figure}

\begin{figure}
    \centering
    \includegraphics[width=\linewidth]{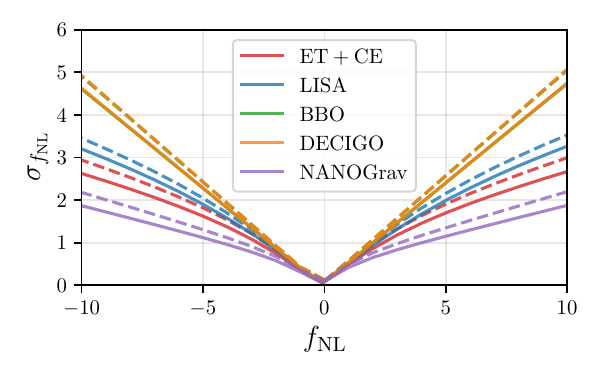}
    \caption{The uncertainties in $f_{\mathrm{NL}}$ are assessed by measuring the auto-correlations (dashed curves) and both auto- and cross-correlations (solid curves) using prospective gravitational-wave detectors. We use $\ell_{\mathrm{max}}=6$ for \ac{NANOGrav} while $\ell_{\mathrm{max}}=30$ for others. For each $f_{\mathrm{NL}}$ value, we calculate $A_{S}$ setting $\mathrm{SNR}=1$ in the auto-correlation-only case, while assuming $g_{\mathrm{NL}}=0$, $\sigma=1$, and $f_{p}=f_{\ast}$. }
    \label{fig:Fig4}
\end{figure}

To assess the performance of detectors in measuring auto- and cross-correlations, we present the \acp{SNR} as a function of $\ell_{\mathrm{max}}$ in Fig.~\ref{fig:Fig3}.
In this analysis, we assume $f_{\mathrm{NL}}=10$, $g_{\mathrm{NL}}=0$, $A_{S}=0.02$, $\sigma=1$, and $f_{p}=f_{\ast}$, where $f_{\ast}$ represents the frequency associated with the optimal sensitivity of detectors.
We examine two scenarios: one focusing solely on auto-correlations and the other involving a combination of auto- and cross-correlations.
Our findings reveal that the inclusion of cross-correlations consistently improves the \ac{SNR} compared to the auto-correlation-only scenario.
Nevertheless, the \ac{SNR} for both cases potentially reaches a saturation point with increasing $\ell_{\mathrm{max}}$ due to the limitations in detectors' angular resolution.
Additionally, we present the \ac{CV} limits for comparison.
For the specified signal, \ac{DECIGO} can achieve the \ac{CV} limits as its noise level is significantly lower than \ac{CV}.
Conversely, other detectors are able to approach the \ac{CV} limits only at the lowest angular multipole moments.

To assess the precision of parameter estimation for the model, we analyze the uncertainties in $f_{\mathrm{NL}}$ measured by the detectors under consideration in Fig.~\ref{fig:Fig4}. 
We use $\ell_{\mathrm{max}}=6$ for \ac{NANOGrav} \cite{Pol:2022sjn} while $\ell_{\mathrm{max}}=30$ for others \cite{Braglia:2021fxn}. 
We examine the same two cases as depicted in Fig.~\ref{fig:Fig3}, but for each $f_{\mathrm{NL}}$ value, we determine $A_{S}$ by setting $\mathrm{SNR}=1$ in the auto-correlation-only case. 
For the specified signal, we observe that the combination of auto- and cross-correlations yields reduced uncertainties in estimating this parameter compared to the case focusing solely on auto-correlations.
By increasing the \ac{SNR}, these uncertainties can be proportionally reduced.

\section{Conclusions and discussion}\label{sec:conclusion}

Our study focused on exploring the cross-correlation between the energy-density anisotropies in \acp{SIGW} and the temperature anisotropies and polarization in the \ac{CMB}.
We derived the expressions for the corresponding angular power spectra and integrated them as template banks into our customized version of \texttt{GW\_CLASS}.
It is worth noting that our analysis accounted for all contributions stemming from the local-type primordial non-Gaussian parameters $f_{\mathrm{NL}}$ and $g_{\mathrm{NL}}$, which have the potential to significantly alter the existing findings in the literature. 
We suggested that the distinct dependencies of the angular power spectra on model parameters could prove instrumental in resolving degeneracies within these parameters. 
Furthermore, we determined the projected sensitivity of multi-band gravitational-wave detectors and \ac{CMB} experiments in detecting these signals and precisely measuring our model parameters, especially those related to primordial non-Gaussianity.
We anticipate that the combination of auto- and cross-correlations will enrich our understanding of the early universe.

The energy-density anisotropies in \acp{SIGW} may have origins partially attributed to the \ac{SW} and \ac{ISW} effects, akin to those observed in the \ac{CMB} but with subtle distinctions.
Gravitons were generated before the radiation-matter equality epoch and traversed longer distances compared to the last scattering of photons.
In contrast to the \ac{CMB}, \acp{SIGW} are additionally influenced by contributions from effective relativistic species and the early \ac{ISW} effect predating the recombination era.
In our present investigation, we considered these effects by utilizing the modified version of \texttt{GW\_CLASS}.
Our findings indicate that the anisotropies in \acp{SIGW} stemming from these effects are at a level of approximately $\sim10^{-4}$, which is comparable to the $\sim10^{-5}$ level observed in the \ac{CMB} \cite{Planck:2018vyg}.
Furthermore, it is important to acknowledge that the \ac{SW} and \ac{ISW} effects are independent of specific models and are inherent in the observations.

The initial inhomogeneities in \acp{SIGW} are inherent to the system and contingent on the specific production mechanism, thus encapsulating information about the underlying model.
In our current study, we consider the non-adiabatic initial conditions that have been extensively explored in our previous works.
Unlike the \ac{CMB}, the significant anisotropies in \acp{SIGW} can be attributed to these initial conditions, particularly in the presence of local-type primordial non-Gaussianity.
In such scenarios, it is anticipated that short-wavelength modes will interact with long-wavelength modes, with the former nonlinearly generating \acp{SIGW}.
Consequently, the spatial distribution of energy density in \acp{SIGW} is influenced by the modulation induced by the latter.

We derived the angular power spectra for the auto- and cross-correlations, revealing their strong dependence on the level of primordial non-Gaussianity.
On the one hand, we may be capable of differentiating \acp{SIGW} from astrophysical foregrounds by examining the distinctive features present in their angular power spectra.
Specifically, the angular power spectra of \acp{SIGW} exhibit characteristic scalings of $\ell(\ell+1)C_{\ell}\sim \ell^{0}$, in contrast to those of \ac{SGWB} originating from \aclp{BBH} which roughly follow $\ell(\ell+1)C_{\ell}\sim\ell^{1}$ \cite{Cusin:2018rsq,Cusin:2017fwz,Cusin:2019jhg,Cusin:2019jpv,Jenkins:2018kxc,Jenkins:2018uac,Jenkins:2019nks,Contaldi:2016koz,Wang:2021djr,Mukherjee:2019oma,Bavera:2021wmw,Bellomo:2021mer}. 
This disparity in scaling suggests the potential to distinguish between these two signal components based on their angular power spectrum characteristics \cite{LISACosmologyWorkingGroup:2022kbp,Cui:2023dlo,Zhao:2024yau}.
On the other hand, our findings indicated that combining the cross-correlation spectra can alleviate the parameter degeneracy observed in the auto-correlation spectrum. 
This outcome has the potential to distinguish between various models of inflation. 
Therefore, our investigation may pave the way for novel approaches to measure non-Gaussian parameters and, consequently, delve into the origins and early evolution of the universe.

We delved deeper into exploring the anticipated sensitivity of forthcoming multi-band gravitational-wave detectors in measuring model parameters, specifically focusing on the non-Gaussian parameter $f_{\mathrm{NL}}$.
Our analysis revealed that the combined consideration of both auto- and cross-correlated angular power spectra could effectively enhance precision in parameter measurements. 
Furthermore, we identified the inherent limitations on the precision of these measurements stemming from \ac{CV}.

{\color{black} We should emphasize that we have assumed scale-invariant local-type primordial non-Gaussianity across the wide range of scales connecting \ac{CMB}/\ac{SIGW} anisotropy (large angular scales) to the much smaller physical scales set by gravitational-wave frequencies. 
Accordingly, we expect that our constraints do not apply to models with strongly scale-dependent $f_{\mathrm{NL}}(k)$, which are left to investigate by future works. }

Our research methodology can be extended to investigate cross-correlations between \acp{SIGW} (or other sources of \ac{SGWB}) and other cosmological probes such as \ac{LSS}, hydrogen 21 cm lines, etc. However, we defer these investigations to future works as they fall outside the scope of our present research.

\begin{acknowledgements}
S.W. and J.P.L. are supported by the National Key R\&D Program of China No. 2023YFC2206403 and the National Natural Science Foundation of China (Grant No. 12175243).  Z.C.Z. is supported by the National Key Research and Development Program of China Grant No. 2021YFC2203001. K.K. is supported by KAKENHI Grant Nos. JP22H05270, JP23KF0289, JP24H01825, JP24K07027. This work is supported by the High-performance Computing Platform of China Agricultural University. 

\end{acknowledgements}

\bibliographystyle{spphys}
\bibliography{apssamp.bib}

\providecommand{\noopsort}[1]{}\providecommand{\singleletter}[1]{#1}%
\begin{thebibliography}{10}
\providecommand{\url}[1]{{#1}}
\providecommand{\urlprefix}{URL }
\expandafter\ifx\csname urlstyle\endcsname\relax
  \providecommand{\doi}[1]{DOI \discretionary{}{}{}#1}\else
  \providecommand{\doi}{DOI \discretionary{}{}{}\begingroup \urlstyle{rm}\Url}\fi

\bibitem{Ananda:2006af}
K.N. Ananda, C.~Clarkson, D.~Wands, Phys. Rev. D \textbf{75}, 123518 (2007).
\newblock \doi{10.1103/PhysRevD.75.123518}

\bibitem{Baumann:2007zm}
D.~Baumann, P.J. Steinhardt, K.~Takahashi, K.~Ichiki, Phys. Rev. D \textbf{76}, 084019 (2007).
\newblock \doi{10.1103/PhysRevD.76.084019}

\bibitem{Espinosa:2018eve}
J.R. Espinosa, D.~Racco, A.~Riotto, JCAP \textbf{09}, 012 (2018).
\newblock \doi{10.1088/1475-7516/2018/09/012}

\bibitem{Kohri:2018awv}
K.~Kohri, T.~Terada, Phys. Rev. D \textbf{97}(12), 123532 (2018).
\newblock \doi{10.1103/PhysRevD.97.123532}

\bibitem{Mollerach:2003nq}
S.~Mollerach, D.~Harari, S.~Matarrese, Phys. Rev. D \textbf{69}, 063002 (2004).
\newblock \doi{10.1103/PhysRevD.69.063002}

\bibitem{Assadullahi:2009jc}
H.~Assadullahi, D.~Wands, Phys. Rev. D \textbf{81}, 023527 (2010).
\newblock \doi{10.1103/PhysRevD.81.023527}

\bibitem{Domenech:2021ztg}
G.~Dom\`enech, Universe \textbf{7}(11), 398 (2021).
\newblock \doi{10.3390/universe7110398}

\bibitem{Maldacena:2002vr}
J.M. Maldacena, JHEP \textbf{05}, 013 (2003).
\newblock \doi{10.1088/1126-6708/2003/05/013}

\bibitem{Bartolo:2004if}
N.~Bartolo, E.~Komatsu, S.~Matarrese, A.~Riotto, Phys. Rept. \textbf{402}, 103 (2004).
\newblock \doi{10.1016/j.physrep.2004.08.022}

\bibitem{Allen:1987vq}
T.J. Allen, B.~Grinstein, M.B. Wise, Phys. Lett. B \textbf{197}, 66 (1987).
\newblock \doi{10.1016/0370-2693(87)90343-1}

\bibitem{Bartolo:2001cw}
N.~Bartolo, S.~Matarrese, A.~Riotto, Phys. Rev. D \textbf{65}, 103505 (2002).
\newblock \doi{10.1103/PhysRevD.65.103505}

\bibitem{Acquaviva:2002ud}
V.~Acquaviva, N.~Bartolo, S.~Matarrese, A.~Riotto, Nucl. Phys. B \textbf{667}, 119 (2003).
\newblock \doi{10.1016/S0550-3213(03)00550-9}

\bibitem{Bernardeau:2002jy}
F.~Bernardeau, J.P. Uzan, Phys. Rev. D \textbf{66}, 103506 (2002).
\newblock \doi{10.1103/PhysRevD.66.103506}

\bibitem{Chen:2006nt}
X.~Chen, M.x. Huang, S.~Kachru, G.~Shiu, JCAP \textbf{01}, 002 (2007).
\newblock \doi{10.1088/1475-7516/2007/01/002}

\bibitem{Byrnes:2010em}
C.T. Byrnes, K.Y. Choi, Adv. Astron. \textbf{2010}, 724525 (2010).
\newblock \doi{10.1155/2010/724525}

\bibitem{Carr:2020xqk}
B.~Carr, F.~Kuhnel, Ann. Rev. Nucl. Part. Sci. \textbf{70}, 355 (2020).
\newblock \doi{10.1146/annurev-nucl-050520-125911}

\bibitem{Carr:2020gox}
B.~Carr, K.~Kohri, Y.~Sendouda, J.~Yokoyama, Rept. Prog. Phys. \textbf{84}(11), 116902 (2021).
\newblock \doi{10.1088/1361-6633/ac1e31}

\bibitem{Carr:2023tpt}
B.~Carr, S.~Clesse, J.~Garcia-Bellido, M.~Hawkins, F.~Kuhnel, Phys. Rept. \textbf{1054}, 1 (2024).
\newblock \doi{10.1016/j.physrep.2023.11.005}

\bibitem{Regimbau:2011rp}
T.~Regimbau, Res. Astron. Astrophys. \textbf{11}, 369 (2011).
\newblock \doi{10.1088/1674-4527/11/4/001}

\bibitem{LISACosmologyWorkingGroup:2022kbp}
N.~Bartolo, et~al., JCAP \textbf{11}, 009 (2022).
\newblock \doi{10.1088/1475-7516/2022/11/009}

\bibitem{Li:2023qua}
J.P. Li, S.~Wang, Z.C. Zhao, K.~Kohri, JCAP \textbf{10}, 056 (2023).
\newblock \doi{10.1088/1475-7516/2023/10/056}

\bibitem{Li:2023xtl}
J.P. Li, S.~Wang, Z.C. Zhao, K.~Kohri, JCAP \textbf{06}, 039 (2024).
\newblock \doi{10.1088/1475-7516/2024/06/039}

\bibitem{Bartolo:2019zvb}
N.~Bartolo, D.~Bertacca, V.~De~Luca, G.~Franciolini, S.~Matarrese, M.~Peloso, A.~Ricciardone, A.~Riotto, G.~Tasinato, JCAP \textbf{02}, 028 (2020).
\newblock \doi{10.1088/1475-7516/2020/02/028}

\bibitem{Wang:2023ost}
S.~Wang, Z.C. Zhao, J.P. Li, Q.H. Zhu, Phys. Rev. Res. \textbf{6}(1), L012060 (2024).
\newblock \doi{10.1103/PhysRevResearch.6.L012060}

\bibitem{Yu:2023jrs}
Y.H. Yu, S.~Wang, Phys. Rev. D \textbf{109}(8), 083501 (2024).
\newblock \doi{10.1103/PhysRevD.109.083501}

\bibitem{Ruiz:2024weh}
J.{\'A}. Ruiz, J.~Rey, JCAP \textbf{04}, 026 (2025).
\newblock \doi{10.1088/1475-7516/2025/04/026}

\bibitem{Rey:2024giu}
J.~Rey,   (2024).
\newblock \doi{10.48550/arXiv.2411.08873}

\bibitem{Li:2024zwx}
J.P. Li, S.~Wang, Z.C. Zhao, K.~Kohri, JCAP \textbf{05}, 109 (2024).
\newblock \doi{10.1088/1475-7516/2024/05/109}

\bibitem{Dimastrogiovanni:2022eir}
E.~Dimastrogiovanni, M.~Fasiello, A.~Malhotra, G.~Tasinato, JCAP \textbf{01}, 018 (2023).
\newblock \doi{10.1088/1475-7516/2023/01/018}

\bibitem{Schulze:2023ich}
F.~Schulze, L.~Valbusa~Dall'Armi, J.~Lesgourgues, A.~Ricciardone, N.~Bartolo, D.~Bertacca, C.~Fidler, S.~Matarrese, JCAP \textbf{10}, 025 (2023).
\newblock \doi{10.1088/1475-7516/2023/10/025}

\bibitem{Cai:2024dya}
R.G. Cai, S.J. Wang, Z.Y. Yuwen, X.X. Zeng, JCAP \textbf{01}, 011 (2025).
\newblock \doi{10.1088/1475-7516/2025/01/011}

\bibitem{Malhotra:2020ket}
A.~Malhotra, E.~Dimastrogiovanni, M.~Fasiello, M.~Shiraishi, JCAP \textbf{03}, 088 (2021).
\newblock \doi{10.1088/1475-7516/2021/03/088}

\bibitem{Perna:2023dgg}
G.~Perna, A.~Ricciardone, D.~Bertacca, S.~Matarrese, JCAP \textbf{10}, 014 (2023).
\newblock \doi{10.1088/1475-7516/2023/10/014}

\bibitem{Adshead:2020bji}
P.~Adshead, N.~Afshordi, E.~Dimastrogiovanni, M.~Fasiello, E.A. Lim, G.~Tasinato, Phys. Rev. D \textbf{103}(2), 023532 (2021).
\newblock \doi{10.1103/PhysRevD.103.023532}

\bibitem{Seljak:1996is}
U.~Seljak, M.~Zaldarriaga, Astrophys. J. \textbf{469}, 437 (1996).
\newblock \doi{10.1086/177793}

\bibitem{Zaldarriaga:1996xe}
M.~Zaldarriaga, U.~Seljak, Phys. Rev. D \textbf{55}, 1830 (1997).
\newblock \doi{10.1103/PhysRevD.55.1830}

\bibitem{Kamionkowski:1996ks}
M.~Kamionkowski, A.~Kosowsky, A.~Stebbins, Phys. Rev. D \textbf{55}, 7368 (1997).
\newblock \doi{10.1103/PhysRevD.55.7368}

\bibitem{Diego_Blas_2011}
D.~Blas, J.~Lesgourgues, T.~Tram, Journal of Cosmology and Astroparticle Physics \textbf{2011}(07), 034–034 (2011).
\newblock \doi{10.1088/1475-7516/2011/07/034}.
\newblock \urlprefix\url{http://dx.doi.org/10.1088/1475-7516/2011/07/034}

\bibitem{Jiang:2022uyg}
J.Q. Jiang, Y.S. Piao, Phys. Rev. D \textbf{105}(10), 103514 (2022).
\newblock \doi{10.1103/PhysRevD.105.103514}

\bibitem{Ye:2022efx}
G.~Ye, J.Q. Jiang, Y.S. Piao, Phys. Rev. D \textbf{106}(10), 103528 (2022).
\newblock \doi{10.1103/PhysRevD.106.103528}

\bibitem{Planck:2018vyg}
N.~Aghanim, et~al., Astron. Astrophys. \textbf{641}, A6 (2020).
\newblock \doi{10.1051/0004-6361/201833910}.
\newblock [Erratum: Astron.Astrophys. 652, C4 (2021)]

\bibitem{Yuan:2023ofl}
C.~Yuan, D.S. Meng, Q.G. Huang, JCAP \textbf{12}, 036 (2023).
\newblock \doi{10.1088/1475-7516/2023/12/036}

\bibitem{Nakama:2016gzw}
T.~Nakama, J.~Silk, M.~Kamionkowski, Phys. Rev. D \textbf{95}(4), 043511 (2017).
\newblock \doi{10.1103/PhysRevD.95.043511}

\bibitem{Garcia-Bellido:2017aan}
J.~Garcia-Bellido, M.~Peloso, C.~Unal, JCAP \textbf{09}, 013 (2017).
\newblock \doi{10.1088/1475-7516/2017/09/013}

\bibitem{Adshead:2021hnm}
P.~Adshead, K.D. Lozanov, Z.J. Weiner, JCAP \textbf{10}, 080 (2021).
\newblock \doi{10.1088/1475-7516/2021/10/080}

\bibitem{Ragavendra:2021qdu}
H.V. Ragavendra, Phys. Rev. D \textbf{105}(6), 063533 (2022).
\newblock \doi{10.1103/PhysRevD.105.063533}

\bibitem{Ragavendra:2020sop}
H.V. Ragavendra, P.~Saha, L.~Sriramkumar, J.~Silk, Phys. Rev. D \textbf{103}(8), 083510 (2021).
\newblock \doi{10.1103/PhysRevD.103.083510}

\bibitem{Garcia-Saenz:2022tzu}
S.~Garcia-Saenz, L.~Pinol, S.~Renaux-Petel, D.~Werth, JCAP \textbf{03}, 057 (2023).
\newblock \doi{10.1088/1475-7516/2023/03/057}

\bibitem{Abe:2022xur}
K.T. Abe, R.~Inui, Y.~Tada, S.~Yokoyama, JCAP \textbf{05}, 044 (2023).
\newblock \doi{10.1088/1475-7516/2023/05/044}

\bibitem{Cai:2018dig}
R.g. Cai, S.~Pi, M.~Sasaki, Phys. Rev. Lett. \textbf{122}(20), 201101 (2019).
\newblock \doi{10.1103/PhysRevLett.122.201101}

\bibitem{Unal:2018yaa}
C.~Unal, Phys. Rev. D \textbf{99}(4), 041301 (2019).
\newblock \doi{10.1103/PhysRevD.99.041301}

\bibitem{Atal:2021jyo}
V.~Atal, G.~Dom\`enech, JCAP \textbf{06}, 001 (2021).
\newblock \doi{10.1088/1475-7516/2021/06/001}

\bibitem{Yuan:2020iwf}
C.~Yuan, Q.G. Huang, Phys. Lett. B \textbf{821}, 136606 (2021).
\newblock \doi{10.1016/j.physletb.2021.136606}

\bibitem{Zhang:2021rqs}
F.~Zhang, Phys. Rev. D \textbf{105}(6), 063539 (2022).
\newblock \doi{10.1103/PhysRevD.105.063539}

\bibitem{Lin:2021vwc}
J.~Lin, S.~Gao, Y.~Gong, Y.~Lu, Z.~Wang, F.~Zhang, Phys. Rev. D \textbf{107}(4), 043517 (2023).
\newblock \doi{10.1103/PhysRevD.107.043517}

\bibitem{Chen:2022dqr}
L.Y. Chen, H.~Yu, P.~Wu, Phys. Rev. D \textbf{106}(6), 063537 (2022).
\newblock \doi{10.1103/PhysRevD.106.063537}

\bibitem{Chang:2023aba}
Z.~Chang, Y.T. Kuang, D.~Wu, J.Z. Zhou, Q.H. Zhu, Phys. Rev. D \textbf{109}(4), L041303 (2024).
\newblock \doi{10.1103/PhysRevD.109.L041303}

\bibitem{Cai:2019elf}
R.G. Cai, S.~Pi, S.J. Wang, X.Y. Yang, JCAP \textbf{10}, 059 (2019).
\newblock \doi{10.1088/1475-7516/2019/10/059}

\bibitem{Perna:2024ehx}
G.~Perna, C.~Testini, A.~Ricciardone, S.~Matarrese, JCAP \textbf{05}, 086 (2024).
\newblock \doi{10.1088/1475-7516/2024/05/086}

\bibitem{Zeng:2024ovg}
X.X. Zeng, R.G. Cai, S.J. Wang, JCAP \textbf{10}, 045 (2024).
\newblock \doi{10.1088/1475-7516/2024/10/045}

\bibitem{ValbusaDallArmi:2020ifo}
L.~Valbusa~Dall'Armi, A.~Ricciardone, N.~Bartolo, D.~Bertacca, S.~Matarrese, Phys. Rev. D \textbf{103}(2), 023522 (2021).
\newblock \doi{10.1103/PhysRevD.103.023522}

\bibitem{Dimastrogiovanni:2021mfs}
E.~Dimastrogiovanni, M.~Fasiello, A.~Malhotra, P.D. Meerburg, G.~Orlando, JCAP \textbf{02}(02), 040 (2022).
\newblock \doi{10.1088/1475-7516/2022/02/040}

\bibitem{LISACosmologyWorkingGroup:2022jok}
P.~Auclair, et~al., Living Rev. Rel. \textbf{26}(1), 5 (2023).
\newblock \doi{10.1007/s41114-023-00045-2}

\bibitem{Unal:2020mts}
C.~\"Unal, E.D. Kovetz, S.P. Patil, Phys. Rev. D \textbf{103}(6), 063519 (2021).
\newblock \doi{10.1103/PhysRevD.103.063519}

\bibitem{Malhotra:2022ply}
A.~Malhotra, E.~Dimastrogiovanni, G.~Dom\`enech, M.~Fasiello, G.~Tasinato, Phys. Rev. D \textbf{107}(10), 103502 (2023).
\newblock \doi{10.1103/PhysRevD.107.103502}

\bibitem{ValbusaDallArmi:2023nqn}
L.~Valbusa~Dall'Armi, A.~Mierna, S.~Matarrese, A.~Ricciardone,   (2023).
\newblock \doi{10.48550/arXiv.2307.11043}

\bibitem{Contaldi:2016koz}
C.R. Contaldi, Phys. Lett. B \textbf{771}, 9 (2017).
\newblock \doi{10.1016/j.physletb.2017.05.020}

\bibitem{Sachs:1967er}
R.K. Sachs, A.M. Wolfe, Astrophys. J. \textbf{147}, 73 (1967).
\newblock \doi{10.1007/s10714-007-0448-9}

\bibitem{Kuwahara:2024jiz}
S.~Kuwahara, L.~Tsukada,   (2024).
\newblock \doi{10.48550/arXiv.2411.19761}

\bibitem{Liang:2024tgn}
Z.C. Liang, Z.Y. Li, E.K. Li, J.d. Zhang, Y.M. Hu, Phys. Rev. D \textbf{111}(4), 043032 (2025).
\newblock \doi{10.1103/PhysRevD.111.043032}

\bibitem{Tian:2024ljg}
C.~Tian, R.~Ding, X.X. Kou, JCAP \textbf{08}, 037 (2025).
\newblock \doi{10.1088/1475-7516/2025/08/037}

\bibitem{ly2017tutorialfisherinformation}
A.~Ly, M.~Marsman, J.~Verhagen, R.~Grasman, E.J. Wagenmakers.
\newblock A tutorial on fisher information (2017).
\newblock \urlprefix\url{https://arxiv.org/abs/1705.01064}

\bibitem{jenet2009northamericannanohertzobservatory}
F.~Jenet, L.S. Finn, J.~Lazio, A.~Lommen, M.~McLaughlin, I.~Stairs, D.~Stinebring, J.~Verbiest, A.~Archibald, Z.~Arzoumanian, D.~Backer, J.~Cordes, P.~Demorest, R.~Ferdman, P.~Freire, M.~Gonzalez, V.~Kaspi, V.~Kondratiev, D.~Lorimer, R.~Lynch, D.~Nice, S.~Ransom, R.~Shannon, X.~Siemens.
\newblock The north american nanohertz observatory for gravitational waves (2009).
\newblock \urlprefix\url{https://arxiv.org/abs/0909.1058}

\bibitem{Baker:2019nia}
J.~Baker, et~al.,   (2019).
\newblock \doi{10.48550/arXiv.1907.06482}

\bibitem{Smith:2019wny}
T.L. Smith, T.L. Smith, R.R. Caldwell, R.~Caldwell, Phys. Rev. D \textbf{100}(10), 104055 (2019).
\newblock \doi{10.1103/PhysRevD.100.104055}.
\newblock [Erratum: Phys.Rev.D 105, 029902 (2022)]

\bibitem{Crowder:2005nr}
J.~Crowder, N.J. Cornish, Phys. Rev. D \textbf{72}, 083005 (2005).
\newblock \doi{10.1103/PhysRevD.72.083005}

\bibitem{Smith:2016jqs}
T.L. Smith, R.~Caldwell, Phys. Rev. D \textbf{95}(4), 044036 (2017).
\newblock \doi{10.1103/PhysRevD.95.044036}

\bibitem{Seto:2001qf}
N.~Seto, S.~Kawamura, T.~Nakamura, Phys. Rev. Lett. \textbf{87}, 221103 (2001).
\newblock \doi{10.1103/PhysRevLett.87.221103}

\bibitem{Kawamura:2020pcg}
S.~Kawamura, et~al., PTEP \textbf{2021}(5), 05A105 (2021).
\newblock \doi{10.1093/ptep/ptab019}

\bibitem{Hild:2008ng}
S.~Hild, S.~Chelkowski, A.~Freise,   (2008).
\newblock \doi{10.48550/arXiv.0810.0604}

\bibitem{Reitze:2019iox}
D.~Reitze, et~al., Bull. Am. Astron. Soc. \textbf{51}(7), 035 (2019).
\newblock \doi{10.48550/arXiv.1907.04833}

\bibitem{Pol:2022sjn}
N.~Pol, S.R. Taylor, J.D. Romano, Astrophys. J. \textbf{940}(2), 173 (2022).
\newblock \doi{10.3847/1538-4357/ac9836}

\bibitem{Braglia:2021fxn}
M.~Braglia, S.~Kuroyanagi, Phys. Rev. D \textbf{104}(12), 123547 (2021).
\newblock \doi{10.1103/PhysRevD.104.123547}

\bibitem{Tegmark:1999ke}
M.~Tegmark, D.J. Eisenstein, W.~Hu, A.~de~Oliveira-Costa, Astrophys. J. \textbf{530}, 133 (2000).
\newblock \doi{10.1086/308348}

\bibitem{Cusin:2018rsq}
G.~Cusin, I.~Dvorkin, C.~Pitrou, J.P. Uzan, Phys. Rev. Lett. \textbf{120}, 231101 (2018).
\newblock \doi{10.1103/PhysRevLett.120.231101}

\bibitem{Cusin:2017fwz}
G.~Cusin, C.~Pitrou, J.P. Uzan, Phys. Rev. \textbf{D96}(10), 103019 (2017).
\newblock \doi{10.1103/PhysRevD.96.103019}

\bibitem{Cusin:2019jhg}
G.~Cusin, I.~Dvorkin, C.~Pitrou, J.P. Uzan, Mon. Not. Roy. Astron. Soc. \textbf{493}(1), L1 (2020).
\newblock \doi{10.1093/mnrasl/slz182}

\bibitem{Cusin:2019jpv}
G.~Cusin, I.~Dvorkin, C.~Pitrou, J.P. Uzan, Phys. Rev. D \textbf{100}(6), 063004 (2019).
\newblock \doi{10.1103/PhysRevD.100.063004}

\bibitem{Jenkins:2018kxc}
A.C. Jenkins, R.~O'Shaughnessy, M.~Sakellariadou, D.~Wysocki, Phys. Rev. Lett. \textbf{122}(11), 111101 (2019).
\newblock \doi{10.1103/PhysRevLett.122.111101}

\bibitem{Jenkins:2018uac}
A.C. Jenkins, M.~Sakellariadou, T.~Regimbau, E.~Slezak, Phys. Rev. D \textbf{98}(6), 063501 (2018).
\newblock \doi{10.1103/PhysRevD.98.063501}

\bibitem{Jenkins:2019nks}
A.C. Jenkins, J.D. Romano, M.~Sakellariadou, Phys. Rev. D \textbf{100}(8), 083501 (2019).
\newblock \doi{10.1103/PhysRevD.100.083501}

\bibitem{Wang:2021djr}
S.~Wang, V.~Vardanyan, K.~Kohri, Phys. Rev. D \textbf{106}(12), 123511 (2022).
\newblock \doi{10.1103/PhysRevD.106.123511}

\bibitem{Mukherjee:2019oma}
S.~Mukherjee, J.~Silk, Mon. Not. Roy. Astron. Soc. \textbf{491}(4), 4690 (2020).
\newblock \doi{10.1093/mnras/stz3226}

\bibitem{Bavera:2021wmw}
S.S. Bavera, G.~Franciolini, G.~Cusin, A.~Riotto, M.~Zevin, T.~Fragos, Astron. Astrophys. \textbf{660}, A26 (2022).
\newblock \doi{10.1051/0004-6361/202142208}

\bibitem{Bellomo:2021mer}
N.~Bellomo, D.~Bertacca, A.C. Jenkins, S.~Matarrese, A.~Raccanelli, T.~Regimbau, A.~Ricciardone, M.~Sakellariadou, JCAP \textbf{06}(06), 030 (2022).
\newblock \doi{10.1088/1475-7516/2022/06/030}

\bibitem{Cui:2023dlo}
Y.~Cui, S.~Kumar, R.~Sundrum, Y.~Tsai, JCAP \textbf{10}, 064 (2023).
\newblock \doi{10.1088/1475-7516/2023/10/064}

\bibitem{Zhao:2024yau}
Z.C. Zhao, S.~Wang, Sci. China Phys. Mech. Astron. \textbf{67}(12), 120411 (2024).
\newblock \doi{10.1007/s11433-024-2498-0}

\end{thebibliography}

\end{document}